\documentclass[aps,pre,twocolumn,showpacs,a4paper]{revtex4}
\usepackage{amsmath,amssymb,amsfonts}
\usepackage{graphicx}

\newcommand{\ham}{\mathcal{H}}
\newcommand{\spc}[1]{\mathsf{#1}}
\newcommand{\phs}[1]{{(\mathrm{#1})}}
\newcommand{\nocc}[2]{{n_{\spc{#1},{#2}}}}
\newcommand{\eref}[1]{(\ref{#1})}

\begin{document}

\title
{
  Inverse swelling of a hydrophobic polymer in aqueous solution
}
\author{M. Pretti}
\affiliation
{
  Consiglio Nazionale delle Ricerche (INFM-CNR),
  Dipartimento di Fisica, \\ Politecnico di Torino,
  Corso Duca degli Abruzzi 24, I-10129 Torino, Italy
}
\date{\today}
\begin{abstract}
We address the problem of inverse polymer swelling. This
phenomenon, in which a collapsed polymer chain swells upon {\em
decreasing} temperature, can be observed experimentally in
so-called thermoreversible homopolymers in aqueous solution, and
is believed to be related to the role of hydrophobicity in protein
folding. We consider a lattice-fluid model of water, defined on a
body-centered cubic lattice, which has been previously shown to
account for most thermodynamic anomalies of water and of
hydrophobic solvation for monomeric solutes. We represent the
polymer as a self-avoiding walk on the same lattice, and
investigate the resulting model at a first order approximation
level, equivalent to the exact calculation on a Husimi lattice.
Depending on interaction parameters and applied pressure, the
model exhibits first and/or second order swelling transitions upon
decreasing temperature.
\end{abstract}

\pacs{
61.25.Hq,   
05.50.+q,   
87.10.+e   
}

\maketitle

\section{Introduction}

From the experimental point of view, water is known to exhibit
several thermodynamic anomalies, both as a pure
substance~\cite{Franks1982,Stanley2003} and as a solvent, in
particular for non-polar (hydrophobic) chemical
species~\cite{BenNaim1980,DillScience1990,SouthallDillHaymet2002}.
The transfer process of a small non-polar solute molecule in water
is characterized by a positive solvation Gibbs free energy (it is
thermodynamically unfavored), a negative solvation enthalpy (it is
energetically favored), a negative solvation entropy (it has an
ordering effect), and a large positive solvation heat
capacity~\cite{BenNaim1987}. More precisely, for prototype
hydrophobic species (that is, for instance, noble gases),
solvation entropies and enthalpies, which are negative at room
temperature, increase upon increasing temperature, and eventually
become positive. Accordingly, the solvation Gibbs free energy
displays a maximum as a function of temperature. These properties
define the so-called hydrophobic effect.

It is quite well established that the hydrophobic effect is an
important driving force for several biophysical
processes~\cite{Tanford1980}, and in particular for protein
folding~\cite{Dill1990}. This is the reason why it has attracted a
high degree of attention in the last years, but a unified
theoretical framework for this phenomenon does not exist yet. It
has been observed that the native folded state of proteins is
maximally stable in the range of temperatures of living organisms,
whereas it tends to be destabilized both by increasing and by
decreasing temperature. For globular proteins, it is eventually
possible to observe complete denaturation also upon decreasing
temperature~\cite{Privalov1990,MakhatadzePrivalov1995}. This
phenomenon is denoted as {\em cold unfolding}, and has a simple
analogue in so-called thermoreversible homopolymers, which exhibit
a transition from a collapsed to a swollen state, upon decreasing
temperature~\cite{Tiktopulo_et_al1995,WuZhou1996,WuWang1998}. We
denote this kind of phase transition as {\em inverse swelling}.
All this phenomenology is qualitatively consistent with the
previously mentioned existence of a maximum in the stabilizing
force, i.e., hydrophobic repulsion, as a function of temperature.
Such reasoning is of course unrigorous, since the hydrophobic
effect is not a real interaction, but an effective repulsion,
resulting from an average over the degrees of freedom of water.

Cold unfolding and inverse swelling have attracted some attention
from the theoretical point of
view~\cite{HansenJensenSneppenZocchi1998,HansenJensenSneppenZocchi1999,BakkHoyeHansenSneppen2001,BakkHansenSneppen2001,DelosriosCaldarelli2000,DelosriosCaldarelli2001,BruscoliniCasetti2000pre,BruscoliniCasetti2001pre,BruscoliniBuzanoPelizzolaPretti2001,BruscoliniBuzanoPelizzolaPretti2002},
and they have been a starting point for highlighting the
importance of water structure details for modeling protein
folding~\cite{SalviDelosrios2003}. One possible way of
investigation consists of computer simulations, based on very
detailed (all-atom) interaction potentials. Unfortunately,
simulations are generally limited by the large computational
effort required. Investigation of a full protein model with
explicit water is still out of reach~\cite{PaschekGarcia2004}, and
also for homopolymers the analysis is generally limited to
relatively short chains or small parameter
regions~\cite{PolsonZuckermann2000,GhoshKalraGarde2005,PaschekNonnGeiger2005}.
A complementary approach involves investigations of simplified
models, either on- or off-lattice. Several models of this kind
have been
proposed~\cite{HansenJensenSneppenZocchi1998,HansenJensenSneppenZocchi1999,BakkHoyeHansenSneppen2001,BakkHansenSneppen2001,DelosriosCaldarelli2000,DelosriosCaldarelli2001,BruscoliniCasetti2000pre,BruscoliniCasetti2001pre,BruscoliniBuzanoPelizzolaPretti2001,BruscoliniBuzanoPelizzolaPretti2002}.
As previously mentioned, it turns out that, in order to reproduce
inverse swelling (or cold unfolding), it is necessary to take into
account, even in extremely simplified fashions, the degrees of
freedom of water. Nevertheless, such descriptions are generally
proposed ad-hoc for this kind of problem. The focus is mainly on
the polymer, whereas water models by themselves would not be
satisfactory models of water.

Conversely, in this article, we start by considering a
lattice-fluid model of water, whose thermodynamic properties have
been previously investigated in
detail~\cite{PrettiBuzano2004,PrettiBuzano2005}. This model
predicts most of the thermodynamic anomalies of real water at
constant pressure (a temperature of maximum density, a minimum of
isothermal compressibility and specific heat), and also a
liquid-liquid phase separation in the supercooled region, and a
second critical point~\cite{PrettiBuzano2004}. Moreover, a
corresponding model for aqueous solutions of ideally inert
monomeric solutes turns out to exhibit the above mentioned
fingerprints of hydrophobicity, and in particular the maximum in
the solvation free energy as a function of
temperature~\cite{PrettiBuzano2005}. We describe the polymer in
solution as a lattice self-avoiding walk, whose steps connect
nearest neighbor sites. We assume that each visited site,
representing a monomeric unit, interacts with water and with other
non-consecutive monomeric units, in the same way as the elementary
solute of the the original model does. In some sense, our model is
unbiased.

The water model is a simplified version of the one proposed by
Roberts and
Debenedetti~\cite{RobertsDebenedetti1996,RobertsPanagiotopoulosDebenedetti1996,RobertsKarayiannakisDebenedetti1998}.
It is defined on the body-centered cubic lattice, and water
molecules possess four equivalent bonding arms arranged in a
tetrahedral symmetry. According to this model, the microscopic
description of water anomalies is based on the competition between
an isotropic (van der Waals-like) interaction and a highly
directional (hydrogen bonding) interaction, and on the difference
between the respective optimal interaction distances. In the
lattice environment, the latter is taken into account by a trick,
that is, a weakening of hydrogen bonds by neighboring water
molecules.

The previously mentioned results on this model have been obtained
by a first order approximation on a tetrahedral cluster
(equivalent to the exact calculation for a Husimi
lattice~\cite{Pretti2003} made up of tetrahedral building blocks),
but turn out to compare quite well with Monte Carlo
simulations~\cite{PrettiBuzano2004}. The same kind of
approximation has been independently verified to be rather
accurate for the case of a semiflexible self-interacting polymer
chain~\cite{Pretti2002} with no explicit solvent. We thus expect
that the calculation presented here, based on the same
approximation technique, could produce reliable results as well,
although it is being applied to a more complicated model involving
water-polymer interactions.

\section{The model}

\begin{figure}[t!]
  \includegraphics*[50mm,186mm][135mm,252mm]{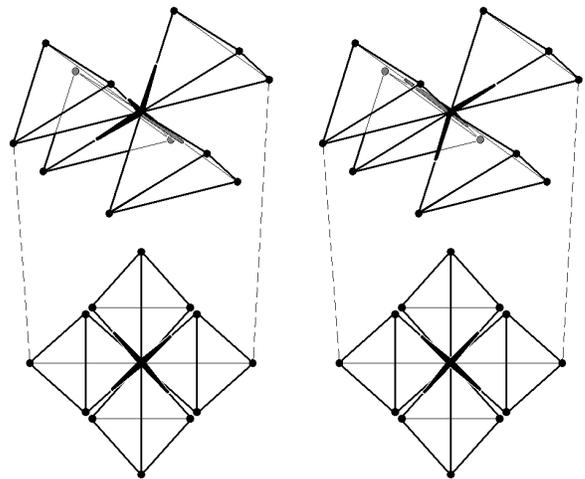}
  \caption
  {
    Husimi lattice structure and bonding water configurations
    $i=1$ (left) and $i=2$ (right).
    Views from top are meant to explain
    configuration symbols used in
    Table~\protect\ref{tab:configurazioni}.
  }
  \label{fig:cacmol}
\end{figure}
Let us now describe the model in detail. As previously mentioned,
it is defined on the bcc lattice. Each site may be empty or
occupied by a water molecule ($\spc{w}$) or by a polymer segment
($\spc{s}$). Since we find it more convenient to work in the
grand-canonical ensemble, each of the two species contributes to
the hamiltonian by a different chemical potential term
$-\mu_{\spc{x}}$, where $\spc{x}\in\{\spc{w},\spc{s}\}$. An
attractive (Van der Waals) energy $-\epsilon_{\spc{x} \spc{y}}<0$
is assigned to any pair of nearest neighbor (NN) sites occupied by
$\spc{x},\spc{y}$ molecules, where $\spc{x}$ and $\spc{y}$ run
over $\{\spc{w},\spc{s}\}$. Of course, the interaction energy
between polymer segments $-\epsilon_{\spc{s} \spc{s}}$ is taken
into account only if the segments are not consecutive along the
chain. Water molecules possess four equivalent arms that can form
hydrogen bonds, arranged in a tetrahedral symmetry, so that they
can point towards 4~out~of~8 NNs of a given site. A hydrogen bond
is formed whenever two NN molecules have a bonding arm pointing to
each other, yielding an energy~$-\eta<0$. Only 2 different water
configurations can form hydrogen bonds (see
Fig.~\ref{fig:cacmol}); $w$~more configurations are allowed, in
which water molecules cannot form bonds (the $w$~parameter is
related to the bond-breaking entropy). An energy increase~$\eta
c_\spc{x}/6$, with $c_\spc{x}\in[0,1]$, is added for each of the
$6$ sites closest to a bond (i.e., 3 out of 6 next nearest
neighbors of each bonded molecule), occupied by an $\spc{x}$
molecule. As far as water molecules are concerned, the weakening
parameter mainly accounts for the fact that hydrogen bonds are
most favorably formed when water molecules are located at a
certain distance, larger than the optimal Van der Waals distance.
Therefore, if too many molecules are present, the average distance
among them decreases, and hydrogen bonds are (on average)
weakened. Moreover, the presence of an external molecule may
perturb the electronic density, resulting in a lowered hydrogen
bond strength as well. A weakening parameter~$c_\spc{s}$ for
polymer segments could take into account possible perturbation
effects for a generic chemical species. Nevertheless, in this
paper we shall consider only the case of an ideal monomer with
$c_\spc{s}=0$.

The bcc lattice model can be conveniently replaced by an analogous
model defined on a Husimi lattice made up of irregular tetrahedral
building blocks (see Fig.~\ref{fig:cacmol}), as explained in
detail in Refs.~\onlinecite{PrettiBuzano2004}
and~\onlinecite{PrettiBuzano2005}. The latter model, which can be
studied exactly, is equivalent to a generalized first order
approximation (on the tetrahedral cluster) for the original
model~\cite{Pretti2003}. From now on, we shall refer to the Husimi
lattice model only. The hamiltonian of the system can be written
as a sum over the tetrahedral clusters
\begin{equation}
  \ham =
  \sum_{\langle \alpha,\beta,\gamma,\delta \rangle}
  \ham_{i_\alpha i_\beta i_\gamma i_\delta}
  ,
  \label{eq:ham}
\end{equation}
where $\ham_{ijkl}$ is the contribution of each cluster, which
will be referred to as tetrahedron hamiltonian, and the subscripts
$i_\alpha,i_\beta,i_\gamma,i_\delta$ label site configurations for
the $4$ vertices ($\alpha,\beta,\gamma,\delta$). It is understood
that the latter are enumerated in a conventional order, for
instance clockwise, with reference to Fig.~\ref{fig:cacmol}
(bottom). Accordingly, $(\alpha,\beta)$, $(\beta,\gamma)$,
$(\gamma,\delta)$, and $(\delta,\alpha)$ turn out to be NN pairs,
whereas $(\alpha,\gamma)$ and $(\beta,\delta)$ are next nearest
neighbors.

Site configurations are reported in
Table~\ref{tab:configurazioni}, where $i=1,2,3$ denote water
molecule configurations (defined as in
Refs.~\onlinecite{PrettiBuzano2004,PrettiBuzano2005}) and
$i=4,\dots,8$ denote local configurations of the polymer chain. As
far as polymer configurations are concerned, let us notice that
they are more conveniently defined with respect to the
tetrahedron, than with respect to a fixed reference frame. This
trick allows us to write a unique form for the tetrahedron
hamiltonian, which otherwise would be orientation-dependent.
Moreover, although in the isotropy hypothesis every local chain
configuration should be equally probable, we need to distinguish
different classes of configurations (each one with a given
multiplicity) in order to impose connectivity constraints, as it
will be clarified below. As a consequence, the Husimi lattice
approximation introduces a small artifact, in that the probability
of the configurations $i=4,6$ (in which the chain locally stays in
the same cluster) turns out to be slightly different from the
probability of the configurations $i=5,7,8$ (in which the chain
passes from one cluster to another). We expect that such artifact
is negligible, according to the results of
Ref.~\onlinecite{Pretti2002}.

The tetrahedron hamiltonian can be written as
\begin{equation}
  \ham_{ijkl} = H_{ijkl} + H_{jkli} + H_{klij} + H_{lijk} +
  L_{ijkl}
  ,
  \label{eq:tetraham1}
\end{equation}
where
\begin{eqnarray}
  H_{ijkl}
  & = &
  - \epsilon_\spc{xy} \nocc{x}{i} \nocc{y}{j}
  \, \overline{b_i^+ b_j^-}
  - \eta h_{ij} \left( 1 - c_\spc{x} \frac{\nocc{x}{k} +
  \nocc{x}{l}}{2}\right)
  \nonumber \\ &&
  - \frac{1}{4} \mu_\spc{x} \nocc{x}{i}
  + I \left( b_i^+ \oplus b_j^- \right)
  .
  \label{eq:tetraham2}
\end{eqnarray}
Let us analyze the various terms appearing in these equations.

The first line of Eq.~\eref{eq:tetraham2} is basically equivalent
to the corresponding term for the mixture model of water and a
simple monomeric solute [see Eq.~(3) in
Ref.~\onlinecite{PrettiBuzano2005}], and includes the Van der
Waals and the hydrogen bonding terms. Occupation variables
$\nocc{x}{i}$ are defined as $\nocc{x}{i}=1$ if the
configuration~$i$ corresponds to a chemical species~$\spc{x}$, and
$\nocc{x}{i}=0$ otherwise (see Table~\ref{tab:configurazioni}),
whereas $h_{ij}$ are hydrogen bond variables, defined as
$h_{ij}=1$ if the pair configuration $(i,j)$ represents a hydrogen
bond (i.e., if $i=1$ and $j=2$), and $h_{ij}=0$ otherwise. It is
understood that repeated $\spc{x}$ and $\spc{y}$ indices are
summed over their possible values $\spc{w},\spc{s}$. The only
difference with respect to the monomeric solute case is related to
the fact that, as previously mentioned, we have to exclude Van der
Waals interactions between consecutive polymer segments. To do so,
we have defined the bond numbers $b_i^+$ (resp. $b_i^-$), which
are boolean variables equal to $1$ if the configuration $i$
represents a polymer segment forming a chemical bond in the
clockwise (resp. counterclockwise) direction of the reference
cluster, and $0$ otherwise (see Table~\ref{tab:configurazioni}).
In this way, the multiplying term $\overline{b_i^+b_j^-}$ (where
the overline denotes boolean negation) is $0$ if $(i,j)$
represents two consecutive segments, and $1$ otherwise.

The second line of Eq.~\eref{eq:tetraham2} includes the chemical
potential contributions (multiplied by $1/4$ to avoid
overcounting) and an infinite energy penalty $I\to\infty$,
assigned to $(i,j)$ configurations which violate connectivity
constraint, i.e., when $i$ wants to form a chemical bond (in the
clockwise direction) and $j$ does not (in the counterclockwise
direction), or vice versa. This is obtained by the exclusive-or
factor $\left( b_i^+ \oplus b_j^- \right)$. Let us notice that the
infinite energy penalty can be treated numerically, since we have
to deal only with Boltzmann factors of the tetrahedron
hamiltonian. Finally, the additive term $L_{ijkl}$ in
Eq.~\eref{eq:tetraham1} is another constraint term, needed to
forbid short loops on tetrahedra. It is therefore simply defined
as $L_{6666}=\infty$ and $L_{ijkl}=0$ otherwise.

\begin{table}
  \caption{
    Site configurations, with corresponding labels ($i$),
    multiplicities~($w_i$),
    occupation variables ($\nocc{w}{i},\nocc{s}{i}$),
    and chemical bond numbers ($b_i^+,b_i^-$).
    Polymer configurations are defined
    with reference to the cluster denoted by a dot;
    $i',i'',i'''$ denote the configurations ``viewed'' by the other $3$ clusters:
    $i=0$ empty site;
    $i=1,2$ bonding water;
    $i=3$ non-bonding water;
    $i=4$ segments in the same cluster, out of the reference one;
    $i=5$ segments in different clusters, out of the reference one;
    $i=6$ segments in the same cluster, the reference one;
    $i=7$ segments in different clusters, one in the reference cluster
    (``clockwise'' direction);
    $i=8$ segments in different clusters, one in the reference cluster
    (``counterclockwise'' direction);
  }
  \begin{ruledtabular}
  \begin{tabular}{l|ccccccccc}
    &
    \resizebox{7mm}{!}{\includegraphics*[55mm,233mm][69mm,247mm]{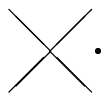}}
    &
    \resizebox{7mm}{!}{\includegraphics*[55mm,233mm][69mm,247mm]{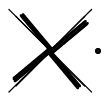}}
    &
    \resizebox{7mm}{!}{\includegraphics*[55mm,233mm][69mm,247mm]{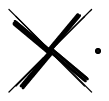}}
    &
    \resizebox{7mm}{!}{\includegraphics*[55mm,233mm][69mm,247mm]{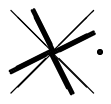}}
    &
    \resizebox{7mm}{!}{\includegraphics*[55mm,233mm][69mm,247mm]{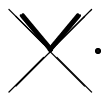}}
    &
    \resizebox{7mm}{!}{\includegraphics*[55mm,233mm][69mm,247mm]{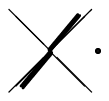}}
    &
    \resizebox{7mm}{!}{\includegraphics*[55mm,233mm][69mm,247mm]{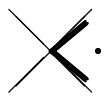}}
    &
    \resizebox{7mm}{!}{\includegraphics*[55mm,233mm][69mm,247mm]{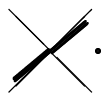}}
    &
    \resizebox{7mm}{!}{\includegraphics*[55mm,233mm][69mm,247mm]{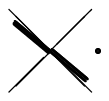}}
    \cr
    \hline
    $i$           & $0$ & $1$ & $2$ & $3$ & $4$ & $5$ & $6$ & $7$ & $8$ \cr
    \hline
    $i'$          & $0$ & $1$ & $2$ & $3$ & $6$ & $5$ & $4$ & $8$ & $7$ \cr
    $i''$         & $0$ & $1$ & $2$ & $3$ & $4$ & $7$ & $4$ & $5$ & $5$ \cr
    $i'''$        & $0$ & $1$ & $2$ & $3$ & $4$ & $8$ & $4$ & $5$ & $5$ \cr
    \hline
    $w_i$         & $1$ & $1$ & $1$ & $w$ & $3$ &$12$ & $1$ & $6$ & $6$ \cr
    \hline
    $\nocc{w}{i}$ & $0$ & $1$ & $1$ & $1$ & $0$ & $0$ & $0$ & $0$ & $0$ \cr
    $\nocc{s}{i}$ & $0$ & $0$ & $0$ & $0$ & $1$ & $1$ & $1$ & $1$ & $1$ \cr
    \hline
    $b_i^+$       & $0$ & $0$ & $0$ & $0$ & $0$ & $0$ & $1$ & $1$ & $0$ \cr
    $b_i^-$       & $0$ & $0$ & $0$ & $0$ & $0$ & $0$ & $1$ & $0$ & $1$
  \end{tabular}
  \end{ruledtabular}
  \label{tab:configurazioni}
\end{table}

\section{The first order (Husimi lattice) approximation}

As mentioned in the Introduction, we perform the exact analysis of
the Husimi lattice model, which is equivalent to the generalized
first order approximation on tetrahedral clusters. The calculation
closely follows the one of the ordinary mixture model, performed
in Ref.~\onlinecite{PrettiBuzano2005}, so that we do not give much
detail. Basically, one has to take into account that the system is
locally tree-like, which allows to write a recursion
equation~\cite{Pretti2003}, for so-called partial partition
functions, defined as follows. Let us assume that the lattice is
actually a tree (Husimi tree), and let us consider just one
branch, with the corresponding partial hamiltonian, obtained by
Eq.~\eqref{eq:ham} with the sum restricted to tetrahedra in the
branch. The partial partition function~$Q_i$ is a sum of Boltzmann
weights of the partial hamiltonian, taken over the configurations
of the branch except the base site (this is why the partial
partition function depends on the base site configuration
variable~$i$). It is convenient to define normalized partial
partition functions $q_i \propto Q_i$, for instance in such a way
that
\begin{equation}
  \sum_{i=0}^8 q_i = 1
  .
\end{equation}
If the branch becomes infinite, i.e., in the thermodynamic limit,
and in the homogeneity hypothesis, the subbranches attached to the
first tetrahedral cluster are equivalent to the main one, so that
one can write the recursion equation
\begin{equation}
  q_i = y^{-1}
  \sum_{j=0}^8 w_j \sum_{k=0}^8 w_k \sum_{l=0}^8 w_l
  e^{-\ham_{ijkl}/T}
  \prod_{\xi = \prime}^{\prime\prime\prime}
  q_{j^\xi} q_{k^\xi} q_{l^\xi}
  ,
  \label{eq:rec}
\end{equation}
where the sum runs over configuration variables in the tetrahedron
except~$i$, $T$ is the temperature expressed in energy units, and
$y$~is a normalization constant.

Let us notice a subtle but important difference with respect to
the corresponding equation for the monomeric solute [Eq.~(8) in
Ref.~\onlinecite{PrettiBuzano2005}]. As mentioned in the previous
section, we have defined local chain configurations with reference
to a given cluster, which allows us to write one single
tetrahedron hamiltonian, independently of orientation. Therefore,
when we consider the operation of ``attaching branches'' to a base
cluster, we have to take into account configurations of sites in
the cluster as they are ``viewed'' by the attached branches. To do
so, we have to write a product over
$\xi=\prime,\prime\prime,\prime\prime\prime$, which denote
precisely the different views, according to
Table~\ref{tab:configurazioni}.

The recursion equation can be iterated numerically to determine a
fixed point. The site configuration probabilities $p_i$ can be
computed by considering the operation of attaching $4$~branches to
the given site, yielding
\begin{equation}
  p_i = z^{-1} q_i
  \prod_{\xi = \prime}^{\prime\prime\prime}
  q_{i^\xi}
  ,
  \label{eq:pjoint}
\end{equation}
where the normalization constant is determined as
\begin{equation}
  z =
  \sum_{i=0}^8
  w_i q_i
  \prod_{\xi = \prime}^{\prime\prime\prime}
  q_{i^\xi}
  .
  \label{eq:zjoint}
\end{equation}
From the knowledge of site configuration probabilities~$p_i$, one
can easily compute the densities, i.e., the
probabilities~$\rho_\spc{w},\rho_\spc{s}$ that a site is occupied
by a water molecule or by a polymer segment, respectively. For
$\spc{x}\in\{\spc{w},\spc{s}\}$, we have
\begin{equation}
  \rho_\spc{x} = \sum_{i=0}^8 w_i p_i \nocc{x}{i}
  ,
  \label{eq:density}
\end{equation}
where the occupation numbers $\nocc{x}{i}$ are explicitly given in
Table~\ref{tab:configurazioni}.

In the presence of multiple fixed points (which can be reached
from different initial conditions), i.e., in the presence of
coexistence phenomena, the first order transition can be
determined by minimizing the (grand-canonical) free energy per
site
\begin{equation}
  \omega = - T \left( \ln y - 2 \ln z \right)
  ,
  \label{eq:granpot}
\end{equation}
where $y$ and $z$ are the normalization constants of the recursion
equation and of the site probability distribution, respectively.
The derivation of this expression requires some manipulations, and
is left to the Appendix. From the knowledge of the free energy,
one can in principle compute all other thermodynamic properties.
Assuming the volume per site equal to~$1$, pressure can be
expressed in energy units as $P = -\omega$.

\section{Results}

First of all, we have to fix a set of model parameters. As far as
water is concerned, we choose the values of our previous
investigation~\cite{PrettiBuzano2004}: $\epsilon_\spc{ww}/\eta =
0.25$, $w = 20$, and $c_\spc{w}=0.5$. The hydrogen bond energy
$\eta$ is taken as the energy unit. The water-water van der Waals
energy $\epsilon_\spc{ww}$ is significantly smaller. The
multiplicity~$w$ of non-bonding water configurations is large, to
mimic the high directionality of hydrogen bonds.

With this set of parameters, the model provides a qualitatively
consistent description of the phase diagram and thermodynamic
anomalies of pure water~\cite{PrettiBuzano2004}, as already
mentioned in the Introduction. For the liquid phase, one observes
a density maximum as a function of temperature at fixed pressure,
i.e., a change of sign in the isobaric thermal expansion
coefficient
\begin{equation}
  \alpha_P = - \frac{\partial \ln \rho_\spc{w}}{\partial T}\biggr\lvert_P
  .
\end{equation}
The temperature of maximum density slightly decreases upon
increasing pressure, as observed in experiments. The isothermal
compressibility
\begin{equation}
  \kappa_T = \frac{\partial \ln \rho_\spc{w}}{\partial P}\biggr\lvert_T
\end{equation}
and the isobaric specific heat exhibit a minimum as a function of
temperature, at constant pressure. Moreover, in the supercooled
regime, the model predicts the so-called second critical point,
which has been conjectured and observed in
simulations~\cite{Stanley2003}, and of which also some
experimental evidences have been found~\cite{MishimaStanley1998}.
The liquid-vapor spinodal displays no reentrance in the positive
pressure half-plane, whereas the temperature of maximum density
locus exhibits a peculiar ``nose-shaped'' reentrance. Let us
recall that this scenario qualitatively agrees with recent
molecular dynamics simulations of water phenomenological
potentials in the negative pressure region~\cite{Stanley2003},
whereas the reentrant spinodal scenario is an older conjecture
invoked to explain thermodynamic anomalies of liquid
water~\cite{Speedy1982I,ZhengDurbenWolfAngell1991}. The
plausibility of the latter has been recently a matter of
debate~\cite{Speedy2004,Debenedetti2004}, but seems to be
definitely ruled out.

\begin{figure}[t!]
  \includegraphics*[42mm,192mm][128mm,245mm]{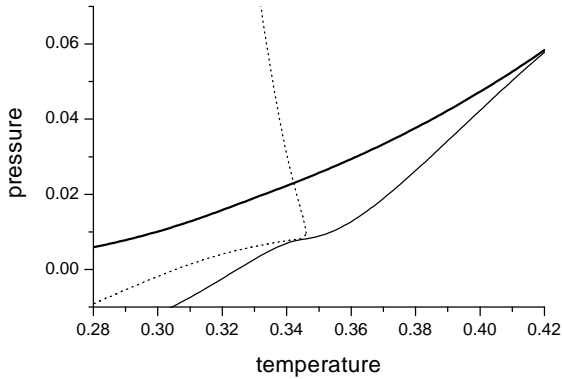}
  \caption
  {
    Pressure ($P/\eta$) vs temperature ($T/\eta$) phase diagram for pure water.
    The solid line denotes the liquid-vapor transition line,
    the thin solid line denotes the liquid phase spinodal,
    whereas the thin dotted line denotes
    the temperature of maximum density locus.
    Parameter values: $\epsilon_{\spc{ww}}/\eta=0.25$, $w=20$, $c_{\spc{w}}=0.5$.
  }
  \label{fig:tp}
\end{figure}

\begin{figure}[t!]
  \includegraphics*[42mm,167mm][128mm,245mm]{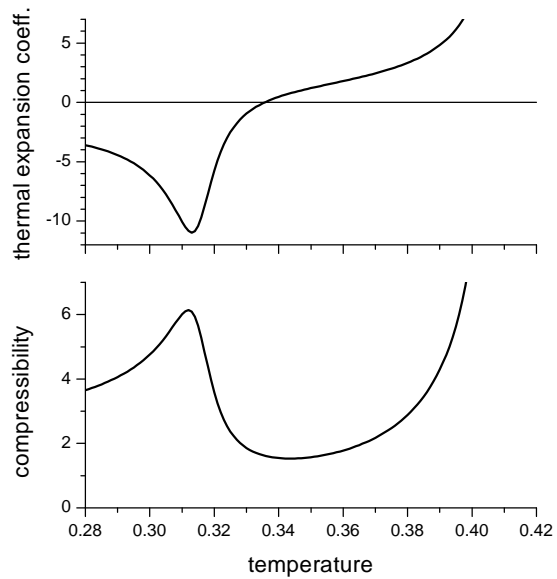}
  \caption
  {
    Pure water response functions at fixed pressure ($P/\eta=0.05$)
    as a function of temperature ($T/\eta$).
    Upper graph: isobaric thermal expansion coefficient ($\eta\alpha_P$).
    Lower graph: isothermal compressibility ($\eta\kappa_T$).
    Parameter values: $\epsilon_{\spc{ww}}/\eta=0.25$, $w=20$, $c_{\spc{w}}=0.5$.
  }
  \label{fig:trespfun}
\end{figure}

We summarize the results obtained by our model for pure water in
Fig.~\ref{fig:tp} (phase diagram) and Fig.~\ref{fig:trespfun}
(response functions), in order to delimit the temperature range in
which we ought to expect the (stable or metastable) liquid water
regime. Let us notice that the response functions correctly
exhibit a divergent behavior, upon approaching the liquid phase
spinodal.

\begin{figure}[t!]
  \includegraphics*[42mm,192mm][128mm,245mm]{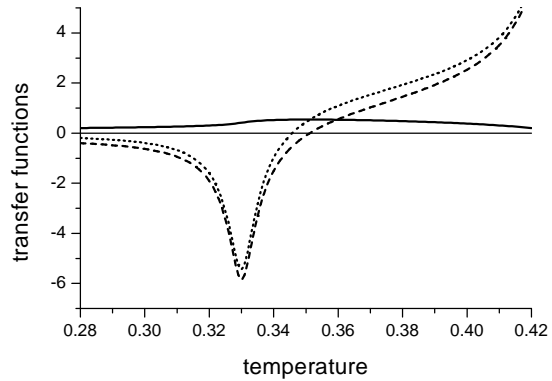}
  \caption
  {
    Transfer energy functions ($E/\eta$) vs temperature ($T/\eta$)
    for a single monomer at liquid-vapor coexistence:
    $E=\Delta g^*$ (solid line),
    $E=T\Delta s^*$ (dashed line),
    $E=\Delta h^*$ (dotted line).
    Parameter values: $\epsilon_{\spc{ww}}/\eta=0.25$, $w=20$, $c_{\spc{w}}=0.5$,
    $\epsilon_{\spc{ws}}=\epsilon_{\spc{ss}}=0$, $c_{\spc{s}}=0$.
  }
  \label{fig:tdfs}
\end{figure}

Upon inserting an ideal inert solute, i.e., adding a monomeric
$\spc{s}$ chemical species, characterized by
$\epsilon_\spc{ws}=\epsilon_\spc{ss}=0$ and $c_\spc{s}=0$, we have
shown that the model is able to reproduce in a qualitatively
accurate way also the solvation properties of simple hydrophobic
solutes, such as noble gases~\cite{PrettiBuzano2005}. Since in
Ref.~\onlinecite{PrettiBuzano2005} we have not considered the
present set of water parameters, we report here the results, in
terms of solvation free energy, entropy, and enthalphy as a
function of temperature (Fig.~\ref{fig:tdfs}).

The solvation Gibbs free energy per molecule $\Delta g^*$, i.e.,
the transfer free energy of a solute molecule from vapor to liquid
phase at coexistence, can be computed as
\begin{equation}
  \Delta g^* = -T \ln
  \frac{\rho_\spc{s}^\phs{l}}{\rho_\spc{s}^\phs{v}}
  ,
  \label{eq:dg}
\end{equation}
where $\rho_\spc{s}^\phs{v}$ and $\rho_\spc{s}^\phs{l}$ denote
solute densities in the two phases, respectively. The solvation
entropy is defined as
\begin{equation}
  \Delta s^* = - \frac{\partial \Delta g^*}
  {\partial T}\biggr\lvert_P
  ,
  \label{eq:ds}
\end{equation}
and the solvation enthalpy can then be computed as
\begin{equation}
  \Delta h^* = \Delta g^* + T \Delta s^*
  .
\end{equation}

The typical experimental temperature range for several hydrophobic
compounds is around $0^\circ\mathrm{C}$ to
$300^\circ\mathrm{C}$~\cite{BenNaim1987}. According to pure water
results, in our model this roughly corresponds to the range
$T/\eta \approx 0.33$ (just below the temperature of maximum
density) to $T/\eta \approx 0.38$ (about half way between the
previous temperature and the critical temperature). In this range,
the solvation Gibbs free energy is positive and displays a broad
maximum, whereas entropy and enthalpy are negative at low
temperatures and positive at high temperatures, in agreement with
experiments~\cite{BenNaim1987}. Upon approaching the critical
point, the free energy tends to zero, whereas the entropy and
enthalphy diverge.

For the polymeric solute, we carry out the investigation as
discussed in the previous Section. We compute, as a composition
variable, the segment molar fraction, defined as
\begin{equation}
  x = \frac{\rho_\spc{s}}{\rho_\spc{w} + \rho_\spc{s}}
  .
\end{equation}
We perform the analysis at constant pressure $P/\eta=0.05$, i.e.,
the same value for which we have computed pure water response
functions. At fixed temperature, we always observe a transition
between a phase with $x=0$ (pure water) at lower values of the
segment chemical potential $\mu_\spc{s}$, and another phase with a
certain positive fraction $x$ of polymer segments ({\em
polymerized phase}) at higher $\mu_\spc{s}$ values. In order to
determine this transition, we have programmed a numerical
procedure, which adjusts the water chemical potential
$\mu_\spc{w}$ in order to fix the pressure of the two phases, and
then finds the $\mu_\spc{s}$ transition value, at which the water
chemical potentials of the two phases are equal. The polymerized
phase at the transition describes the behavior of an isolated
polymer chain in solution~\cite{Vanderzande1998}. Accordingly, the
$x$~value at this point is an indicator of the polymer chain
compactness, as a function of temperature. If $x=0$, the
transition is continuous, and the polymer is in a completely
swollen state, whereas, if $x>0$, the transition is first order,
and the polymer is in a more or less collapsed state.

\begin{figure}[t!]
  \includegraphics*[42mm,192mm][128mm,245mm]{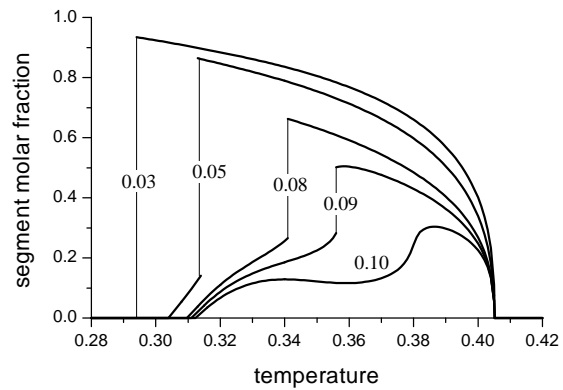}
  \caption
  {
    Polymer compactness
    (segment molar fraction $x$ for the polymerized phase
    at the transition with the pure water phase)
    as a function of temperature ($T/\eta$)
    at fixed pressure ($P/\eta=0.05$),
    for different values of $\epsilon_{\spc{ws}}/\eta$
    (reported on each plot).
    Other parameter values:
    $\epsilon_{\spc{ww}}/\eta=0.25$, $w=20$, $c_{\spc{w}}=0.5$,
    $\epsilon_{\spc{ss}}=0$, $c_{\spc{s}}=0$.
  }
  \label{fig:txs1}
\end{figure}

\begin{figure}[t!]
  \includegraphics*[42mm,132mm][128mm,245mm]{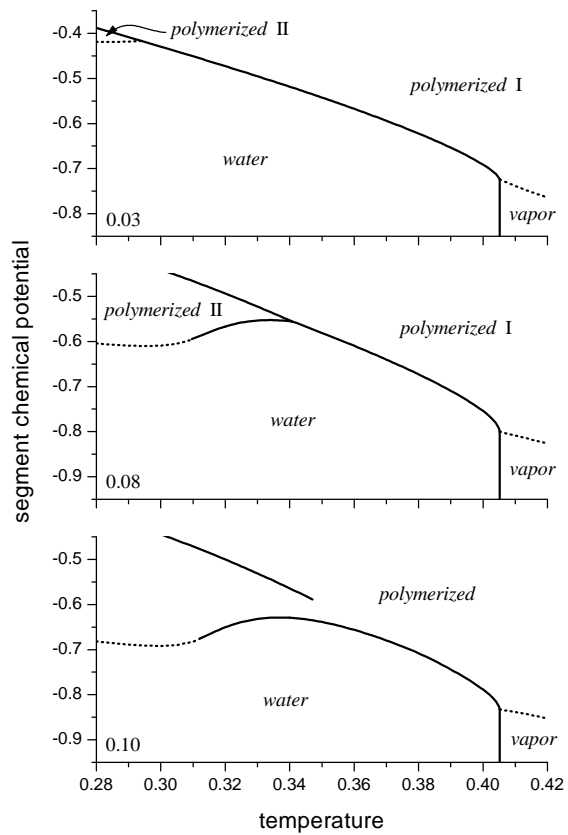}
  \caption
  {
    Segment chemical potential ($\mu_{\spc{s}}/\eta$) vs temperature ($T/\eta$)
    phase diagrams at fixed pressure ($P/\eta=0.05$),
    for different values of $\epsilon_{\spc{ws}}/\eta$
    (reported on each graph).
    Other parameter values:
    $\epsilon_{\spc{ww}}/\eta=0.25$, $w=20$, $c_{\spc{w}}=0.5$,
    $\epsilon_{\spc{ss}}=0$, $c_{\spc{s}}=0$.
    Solid and dashed lines denote first and second order
    transitions, respectively.
  }
  \label{fig:tmus}
\end{figure}

As far as interaction parameters are concerned, we have first
investigated a completely inert (hydrophobic) polymer, assuming
$\epsilon_\spc{ws}=\epsilon_\spc{ss}=0$ and $c_\spc{s}=0$, and
then we have turned on a slight attractive water-segment
interaction ($\epsilon_\spc{ws}>0$). Depending on the value of
this parameter, we have obtained different interesting behaviors.
The results are summarized in Figs.~\ref{fig:txs1}
and~\ref{fig:tmus}.

From $\epsilon_\spc{ws}=0$ to $\epsilon_\spc{ws}/\eta \approx
0.04$, we observe a first-order inverse swelling, since the
segment molar fraction~$x$ abruptly jumps from a high value down
to zero (Fig.~\ref{fig:txs1}). In the $\mu_\spc{s}$ vs $T$ phase
diagram (Fig.~\ref{fig:tmus}, top panel), this transition
corresponds to a critical end-point, which lies among two
different polymerized phases with different densities, and the
water phase. Upon decreasing the segment chemical
potential~$\mu_\spc{s}$, the {\em low} temperature polymerized
phase undergoes a {\em second} order transition to the pure water
phase, corresponding to the {\em swollen} state. On the contrary,
the {\em high} temperature polymerized phase undergoes a {\em
first} order transition to the pure water phase, corresponding to
the {\em collapsed} state. For $\epsilon_\spc{ws}=0$, the inverse
swelling transition occurs around $T/\eta \approx 0.227$. At such
a low temperature, we expect that the liquid water phase predicted
by our model is actually unstable, so that the observed phase
behavior is to be viewed as an
extrapolation~\cite{PrettiBuzano2004}. Upon increasing
$\epsilon_\spc{ws}$, the inverse swelling transition temperature
increases. It is interesting to notice that also an ``ordinary''
swelling transition can be observed at high temperature. Such
transition coincides with the liquid-vapor transition of pure
water, and it is therefore completely unaffected by the
water-segment interaction parameter~$\epsilon_\spc{ws}$. In our
opinion, such feature is a clear evidence that the polymer
collapse is induced by the presence of a dense (liquid) aqueous
solvent around it, i.e., by the hydrophobic effect.

For $\epsilon_\spc{ws}/\eta \gtrsim 0.04$, the behavior changes,
in that inverse swelling occurs via an intermediate, moderately
collapsed phase (Fig.~\ref{fig:txs1}). Upon decreasing
temperature, we first observe a discontinuous jump of the segment
molar fraction~$x$ from a high value to a smaller value, and
subsequently a continuous transition to a completely swollen state
with $x=0$. In the $\mu_\spc{s}$ vs $T$ phase diagram
(Fig.~\ref{fig:tmus}, middle panel), the former transition
corresponds to a triple point among the two polymerized phases and
the water phase. The latter transition corresponds to a
``tricritical'' point, at which the transition between the low
temperature polymerized phases and the pure water phase changes
from first to second order. Such behavior resembles a
$\Theta$~point~\cite{Vanderzande1998} in which the role of
temperature is reversed. Upon increasing $\epsilon_\spc{ws}$, the
temperatures of both transitions increase, and the amplitude of
the discontinuous one is progressively reduced.

Around $\epsilon_\spc{ws}/\eta \approx 0.10$, we observe a totally
continuous inverse swelling, but a reminiscence of the first order
jump can still be observed (Fig.~\ref{fig:txs1}). Such a behavior
corresponds, in the $\mu_\spc{s}$ vs $T$ phase diagram
(Fig.~\ref{fig:tmus}, bottom panel), to the fact that the
transition between the two different polymerized phases ends in a
critical point before encountering the transition line with the
pure water phase. For $\epsilon_\spc{ws}/\eta \approx 0.11$ and
above, we can no longer observe a collapsed phase, and the polymer
is completely swollen at all temperatures. Accordingly, in the
$\mu_\spc{s}$ vs $T$ phase diagram, the transition between the
polymerized phase and the pure water phase is second order at all
temperatures.

\begin{figure}[t!]
  \includegraphics*[42mm,163mm][128mm,245mm]{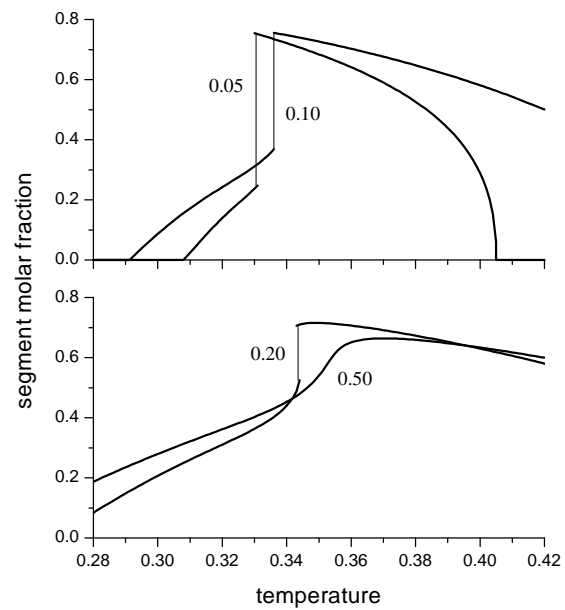}
  \caption
  {
    Polymer compactness
    (segment molar fraction $x$ for the polymerized phase
    at the transition with the pure water phase)
    as a function of temperature ($T/\eta$),
    for different pressure values $P/\eta$ (reported on each plot).
    Parameter values:
    $\epsilon_{\spc{ww}}/\eta=0.25$, $w=20$, $c_{\spc{w}}=0.5$,
    $\epsilon_{\spc{ws}}/\eta=0.07$, $\epsilon_{\spc{ss}}=0$, $c_{\spc{s}}=0$.
  }
  \label{fig:txs2}
\end{figure}

It is noticeable that the effect of pressure on the polymer
behavior is somehow similar to the effect of the water-segment
attractive energy $\epsilon_\spc{ws}$, as shown in
Fig.~\ref{fig:txs2}. The first-order swelling transition
temperature increases upon increasing pressure. The transition
becomes less and less abrupt, and is finally smoothed out at high
pressure values, i.e., in the pressure vs temperature phase
diagram, the transition line ends in a critical point. Let us
notice, by the way, that a similar critical behavior has been
predicted by a simplified model of cold unfolding for
proteins~\cite{HansenJensenSneppenZocchi1999}.

We can also observe some differences. For instance, the high
temperature swelling transition, which was unaffected by the
water-segment interaction at fixed pressure, increases upon
increasing pressure, although, at pressure values lower than the
critical one, it still coincides with the liquid-vapor transition.
This means that indeed, at high temperature, pressure tends to
make the polymer coil more compact. The same holds true for the
low temperature, moderately collapsed phase. This effect is
reversed at very high pressures.

It is important to stress that the different polymer behaviors
described above partially occur in a temperature region which is
likely to be unreachable by experiments. According to
Fig.~\ref{fig:trespfun}, the response functions of pure water
display peaks, but in actual experiments only their high
temperature ``sides'' can be observed. The real peaks are believed
to lie below the homogeneous nucleation
temperature~\cite{Stanley2003}. By the way, this is the reason why
there has been a debate about the possibility of a reentrant
spinodal for liquid water (in the latter case the peaks would
become actual divergences). As far as the polymer problem is
concerned, we conclude that only temperatures above $T/\eta
\approx 0.32$ can be experimentally investigated. We shall return
to this issue in the following.

\begin{figure}[t!]
  \includegraphics*[42mm,192mm][128mm,245mm]{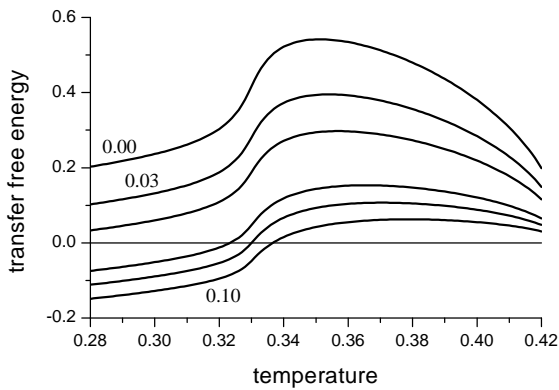}
  \caption
  {
    Transfer free energy ($\Delta g^*/\eta$)
    as a function of temperature ($T/\eta$)
    for a single monomer at liquid-vapor coexistence,
    for $\epsilon_{\spc{ws}}/\eta=0,0.03,0.05,0.08,0.09,0.10$.
    Other parameter values:
    $\epsilon_{\spc{ww}}/\eta=0.25$, $w=20$, $c_{\spc{w}}=0.5$,
    $\epsilon_{\spc{ss}}=0$, $c_{\spc{s}}=0$.
  }
  \label{fig:tdgs}
\end{figure}

As previously mentioned, the present model provides a
qualitatively good description of hydrophobic solvation
thermodynamics for a simple monomeric solute. In particular, if
the monomer is ideally non-interacting, the transfer free energy
displays a maximum as a function of temperature (see
Fig.~\ref{fig:tdfs}). This is expected to be a key ingredient for
inverse swelling of a hydrophobic polymer. Nevertheless, in the
previous investigation, we have taken into account the effect of a
weak water-monomer interaction. Such assumption allows us to
observe inverse swelling at experimentally accessible
temperatures. For completeness, we then compute the solvation free
energies for the monomeric solute, for all the employed values
of~$\epsilon_\spc{ws}/\eta$. The results are reported in
Fig.~\ref{fig:tdgs}.

As a general trend, the curves are shifted towards lower energy
values, upon increasing~$\epsilon_\spc{ws}$, while the maximum
becomes broader. Such behavior qualitatively accounts for
progressive hindering of the collapsed state observed in the
polymer model. Nevertheless, we find it remarkable that, from the
solvation free energy alone, one would not at all expect the
previously described variety of behaviors. In our opinion, this is
a striking evidence of the fact that inverse swelling cannot be
satisfactorily modeled by effective interaction potentials, meant
to take into account a temperature-dependent ``quality of the
solvent'', but it is rather to be ascribed to a subtler interplay
between polymer and water degrees of freedom.

\section{Conclusions}

In this paper, we have proposed and investigated a lattice model
of a linear hydrophobic homopolymer in aqueous solution. As far as
water is concerned, the model is a simplified version of a model
proposed by Roberts and Debenedetti~\cite{RobertsDebenedetti1996},
in which we have removed distinction between donors and acceptors,
which is generally believed not to play a crucial role in the
thermodynamics of hydrogen bonding~\cite{SouthallDillHaymet2002}.
With respect to other simplified model addressed to study the same
kind of systems and phenomenology, the main new ingredient of the
present work is that we have extended a previously studied model
of water and aqueous solutions to the case in which the solute is
a polymer, without other ad-hoc assumptions.

We observe the inverse swelling phenomenon, i.e., a transition
from a collapsed to a swollen state upon decreasing temperature.
For an ideally inert (hydrophobic) polymer, such transition is
first order, and occurs at very low temperatures, where we expect
that liquid water is unstable. In this respect, our model results
can be viewed just as an extrapolation. Nevertheless, the
transition can be moved into the temperature region of liquid
water by an attractive water-monomer interaction. In these
conditions, inverse swelling is no longer purely first order, but
occurs via a discontinuous transition to a moderately collapsed
phase, followed by a $\Theta$-like transition to a completely
swollen state. The latter transition likely stays in the
unobservable region, so that we expect that only first order
swelling could be observed. Upon increasing pressure, the first
order transition temperature increases, the transition is
smoothed, and eventually disappears at high pressure values. Let
us briefly discuss the relevance of such findings to experimental
results.

First of all, the typical system in which the inverse swelling
phenomenon can be observed is an aqueous solution of a
thermoreversible homopolymer, such as poly-N-isopropylacrylamide
(PNIPAM)~\cite{Tiktopulo_et_al1995,WuZhou1996,WuWang1998}. It is
interesting to notice that, for PNIPAM, inverse swelling occurs
actually via a first order transition at a temperature close to
$27^\circ\,\mathrm{C}$, at ordinary pressures. In the framework of
our model, a similar behavior can be reproduced, as previously
mentioned, only in the presence of a water-segment attractive
interaction. For PNIPAM, the existence of such attraction can be
justified by the fact that the monomeric unit (NIPAM) contains in
the side chain a polar (hydrophilic) peptide group, beside ten
non-polar (hydrophobic) hydrogens (the structure is reported for
instance in Ref.~\onlinecite{Tiktopulo_et_al1995}). It is
noteworthy that the assumption of such kind of interaction had to
be made as well in a previous phenomenological treatment of the
inverse swelling phenomenon for
PNIPAM~\cite{BruscoliniBuzanoPelizzolaPretti2002}. In the cited
article, the solvation free energy for NIPAM was assumed as an
effective temperature-dependent interaction with water, and, as a
consequence, inverse swelling turned out to be a $\Theta$-like
point, i.e., a continuous transition. Conversely, from our
investigation it clearly turns out that the information contained
in the solvation free energy is insufficient to discriminate the
type of transition.

A purely hydrophobic short polymer chain in water has been
recently investigated in great detail, by means of molecular
dynamics simulations~\cite{PaschekNonnGeiger2005}. Beside several
structural results, it has turned out that inverse swelling cannot
be observed at ordinary pressures, but only at very high
pressures. The transition is abrupt enough to suppose that it
becomes discontinuous in the thermodynamic limit, and the
transition temperature increases with pressure. These results
qualitatively agree with our findings, although for our model the
effect of pressure is not sufficient to move the transition in the
observable region, for the completely inert polymer.

We have mentioned in the Introduction that some interest on the
inverse swelling phenomenon has been attracted by the fact that
globular proteins undergo a somehow similar phenomenon, known as
cold unfolding. Our model is definitely too simple to describe the
physics of proteins, first of all since a protein is a
heteropolymer, in which the monomeric units have different degrees
of affinity with water. Nonetheless, some interesting analogies
exist. In most cases, the cold unfolding transition of proteins
can be observed in the temperature regime of stable liquid water,
in the presence of denaturants (urea)~\cite{Privalov1990}, or by
applying high pressures (pressure
denaturation)~\cite{Panick_et_al1999,HerberholdWinter2002}.
Extrapolations to zero denaturant concentration, or to ambient
pressure, suggest a transition temperature below the freezing
temperature of water. Both effects seem to be somehow consistent
with our results. On the one hand, a denaturant could be roughly
viewed as something which on average makes the protein more
hydrophilic, i.e., a water-segment attractive interaction. On the
other hand, we have observed that pressure tends to increase the
transition temperature, although this effect is relatively weak.

Let us finally recall that the results presented here have been
obtained by an approximate technique, i.e., a generalized
first-order approximation. Such kind of calculation requires quite
a negligible computational effort, but usually provides
qualitatively reliable results. In particular, the accuracy of the
approximation has been previously verified, by a comparison with
Monte Carlo simulations, both for the pure water
model~\cite{PrettiBuzano2004} and, independently, for an ordinary
self-avoiding walk model with stiffness~\cite{Pretti2002}.
Nevertheless, a similar test for the full model would be welcome,
but this goes beyond the scope of the present paper, and is left
to future work.

\appendix*
\section{Equilibrium free energy}
Hereafter, we justify the formula employed in the text to evaluate
the equilibrium (grand-canonical) free energy density. It is known
that the exact solution of a Husimi lattice can be obtained also
in a variational approach, as a minimum of a generalized Bethe
free energy~\cite{Pretti2003}. Eq.~\eref{eq:granpot} could be
written in principle from the general free energy formula reported
in Ref.~\onlinecite{Pretti2003}, but this formula has been proved
only for the case of scalar configuration variables. Here,
although with some tricks we can label configurations with scalar
quantities, polymer configurations actually behaves as vector
variables, which is revealed by the presence of ``primed''
configurations in Eqs.~\eref{eq:rec} and~\eref{eq:pjoint}.
Therefore, we have to extend the proof, restarting from the
variational Bethe free energy. For our particular system, in the
homogeneity hypothesis, the Bethe free energy per site can be
written as
\begin{equation}
  \omega/T =
  \left\langle \ham_{ijkl}/T + \ln P_{ijkl} \right\rangle
  - 3 \left\langle \ln p_i \right\rangle
  ,
  \label{eq:enlibgen}
\end{equation}
where $\langle\cdot\rangle$~denotes an ensemble average over
configuration variables, and $P_{ijkl}$~is the tetrahedron
probability distribution. Let us notice that the previous formula
holds for a generic free energy density, not necessarily the
grand-canonical one. The tetrahedron probability distribution can
be written as a function of the partial partition functions in the
following form
\begin{equation}
  P_{ijkl} = Z^{-1}
  e^{-\ham_{ijkl}/T}
  \prod_{\xi=\prime}^{\prime\prime\prime}
  q_{i^\xi} q_{j^\xi} q_{k^\xi} q_{l^\xi}
  ,
  \label{eq:pmain}
\end{equation}
which results by considering the operation of attaching $3$
branches to each site of the given tetrahedron. The normalization
constant is
\begin{equation}
  Z =
  \sum_{i=0}^8 w_i \sum_{j=0}^8 w_j \sum_{k=0}^8 w_k \sum_{l=0}^8 w_l
  e^{-\ham_{ijkl}/T}
  \prod_{\xi=\prime}^{\prime\prime\prime}
  q_{i^\xi} q_{j^\xi} q_{k^\xi} q_{l^\xi}
  .
  \label{eq:zmain}
\end{equation}
Replacing Eqs.~\eref{eq:pmain} and~\eref{eq:pjoint}
into~\eref{eq:enlibgen}, and taking into account the linearity of
the average operation, one obtains by simple algebra
\begin{equation}
  \omega/T = - \ln Z + 3 \ln z + \Phi
  ,
  \label{eq:enlib}
\end{equation}
where
\begin{equation}
  \Phi = \sum_{\xi=\prime}^{\prime\prime\prime}
  \left\langle
  \ln \frac{q_{i^\xi}}{q_i}
  \right\rangle =
  \sum_{\xi=\prime}^{\prime\prime\prime}
  \sum_{i=0}^8 w_i p_i
  \left( \ln q_{i^\xi} - \ln q_i \right)
  .
\end{equation}
Expanding the sum, and inserting the multiplicity values $w_i$
reported in Table~\ref{tab:configurazioni}, one can write
\begin{eqnarray}
  \Phi
  & = &
    3  \left(  p_6 - p_4        \right) \ln q_4
  + 3  \left(  p_4 - p_6        \right) \ln q_6
  \nonumber \\ &&
  + 12 \left(  p_7 + p_8 - 2p_5 \right) \ln q_5
  \\ &&
  + 6  \left( 2p_5 + p_8 - 3p_7 \right) \ln q_7
  + 6  \left( 2p_5 + p_7 - 3p_8 \right) \ln q_8
  \nonumber
  ,
\end{eqnarray}
where the terms associated to configurations $i=0,1,2,3$ (empty
site or water) disappear, because in this case $i=i'=i''=i'''$
holds (see Table~\ref{tab:configurazioni}). As mentioned in the
text, one can observe that $p_4=p_6$ and $p_5=p_7=p_8$, whence
$\Phi = 0$. Moreover, making use of
Eqs.~\eref{eq:zmain},~\eref{eq:rec}, and~\eref{eq:zjoint}, it is
possible to show that
\begin{equation}
  Z = y z  ,
\end{equation}
where $y$~is the normalization constant of the recursion
equation~\eref{eq:rec}, which has to be computed at each
iteration, and $z$ is given by Eq.~\eref{eq:zjoint}. Replacing the
previous equation into Eq.~\eref{eq:enlib} with $\Phi=0$, one
finally obtains Eq.~\eref{eq:granpot}, which we aimed to prove.

\acknowledgements

I express my thanks to Carla Buzano for valuable suggestions.


\begin{thebibliography}{43}
\expandafter\ifx\csname natexlab\endcsname\relax\def\natexlab#1{#1}\fi
\expandafter\ifx\csname bibnamefont\endcsname\relax
  \def\bibnamefont#1{#1}\fi
\expandafter\ifx\csname bibfnamefont\endcsname\relax
  \def\bibfnamefont#1{#1}\fi
\expandafter\ifx\csname citenamefont\endcsname\relax
  \def\citenamefont#1{#1}\fi
\expandafter\ifx\csname url\endcsname\relax
  \def\url#1{\texttt{#1}}\fi
\expandafter\ifx\csname urlprefix\endcsname\relax\def\urlprefix{URL }\fi
\providecommand{\bibinfo}[2]{#2}
\providecommand{\eprint}[2][]{\url{#2}}

\bibitem[{\citenamefont{Franks}(1982)}]{Franks1982}
\bibinfo{editor}{\bibfnamefont{F.}~\bibnamefont{Franks}}, ed.,
  \emph{\bibinfo{title}{Water: a Comprehensive Treatise}}
  (\bibinfo{publisher}{Plenum Press, New York}, \bibinfo{year}{1982}).

\bibitem[{\citenamefont{Stanley et~al.}(2003)\citenamefont{Stanley, Buldyrev,
  Giovambattista, Nave, Mossa, Scala, Sciortino, Starr, and
  Yamada}}]{Stanley2003}
\bibinfo{author}{\bibfnamefont{H.~E.} \bibnamefont{Stanley}},
  \bibinfo{author}{\bibfnamefont{S.~V.} \bibnamefont{Buldyrev}},
  \bibinfo{author}{\bibfnamefont{N.}~\bibnamefont{Giovambattista}},
  \bibinfo{author}{\bibfnamefont{E.~L.} \bibnamefont{Nave}},
  \bibinfo{author}{\bibfnamefont{S.}~\bibnamefont{Mossa}},
  \bibinfo{author}{\bibfnamefont{A.}~\bibnamefont{Scala}},
  \bibinfo{author}{\bibfnamefont{F.}~\bibnamefont{Sciortino}},
  \bibinfo{author}{\bibfnamefont{F.~W.} \bibnamefont{Starr}}, \bibnamefont{and}
  \bibinfo{author}{\bibfnamefont{M.}~\bibnamefont{Yamada}},
  \bibinfo{journal}{J. Stat. Phys.} \textbf{\bibinfo{volume}{110}},
  \bibinfo{pages}{1039} (\bibinfo{year}{2003}).

\bibitem[{\citenamefont{Ben-Naim}(1980)}]{BenNaim1980}
\bibinfo{author}{\bibfnamefont{A.}~\bibnamefont{Ben-Naim}},
  \emph{\bibinfo{title}{Hydrophobic interactions}} (\bibinfo{publisher}{Plenum
  Press, New York}, \bibinfo{year}{1980}).

\bibitem[{\citenamefont{Dill}(1990{\natexlab{a}})}]{DillScience1990}
\bibinfo{author}{\bibfnamefont{K.~A.} \bibnamefont{Dill}},
  \bibinfo{journal}{Science} \textbf{\bibinfo{volume}{250}},
  \bibinfo{pages}{297} (\bibinfo{year}{1990}{\natexlab{a}}).

\bibitem[{\citenamefont{Southall et~al.}(2002)\citenamefont{Southall, Dill, and
  Haymet}}]{SouthallDillHaymet2002}
\bibinfo{author}{\bibfnamefont{N.~T.} \bibnamefont{Southall}},
  \bibinfo{author}{\bibfnamefont{K.~A.} \bibnamefont{Dill}}, \bibnamefont{and}
  \bibinfo{author}{\bibfnamefont{A.~D.~J.} \bibnamefont{Haymet}},
  \bibinfo{journal}{J. Phys. Chem. B} \textbf{\bibinfo{volume}{106}},
  \bibinfo{pages}{521} (\bibinfo{year}{2002}).

\bibitem[{\citenamefont{Ben-Naim}(1987)}]{BenNaim1987}
\bibinfo{author}{\bibfnamefont{A.}~\bibnamefont{Ben-Naim}},
  \emph{\bibinfo{title}{Solvation Thermodynamics}} (\bibinfo{publisher}{Plenum
  Press, New York}, \bibinfo{year}{1987}).

\bibitem[{\citenamefont{Tanford}(1980)}]{Tanford1980}
\bibinfo{author}{\bibfnamefont{C.}~\bibnamefont{Tanford}},
  \emph{\bibinfo{title}{The Hydrophobic Effect: Formation of Micelles and
  Biological Membranes}} (\bibinfo{publisher}{Wiley, New York)},
  \bibinfo{year}{1980}), \bibinfo{edition}{2nd} ed.

\bibitem[{\citenamefont{Dill}(1990{\natexlab{b}})}]{Dill1990}
\bibinfo{author}{\bibfnamefont{K.~A.} \bibnamefont{Dill}},
  \bibinfo{journal}{Biochemistry} \textbf{\bibinfo{volume}{29}},
  \bibinfo{pages}{7133} (\bibinfo{year}{1990}{\natexlab{b}}).

\bibitem[{\citenamefont{Privalov}(1990)}]{Privalov1990}
\bibinfo{author}{\bibfnamefont{P.~L.} \bibnamefont{Privalov}},
  \bibinfo{journal}{Crit. Rev. Biochem. Mol. Biol.}
  \textbf{\bibinfo{volume}{25}}, \bibinfo{pages}{281} (\bibinfo{year}{1990}).

\bibitem[{\citenamefont{Makhatadze and
  Privalov}(1995)}]{MakhatadzePrivalov1995}
\bibinfo{author}{\bibfnamefont{G.~I.} \bibnamefont{Makhatadze}}
  \bibnamefont{and} \bibinfo{author}{\bibfnamefont{P.~L.}
  \bibnamefont{Privalov}}, \bibinfo{journal}{Adv. Protein Chem.}
  \textbf{\bibinfo{volume}{47}}, \bibinfo{pages}{307} (\bibinfo{year}{1995}).

\bibitem[{\citenamefont{Tiktopulo et~al.}(1995)\citenamefont{Tiktopulo,
  Uversky, Lushchik, Klenin, Bychkova, and Ptitsyn}}]{Tiktopulo_et_al1995}
\bibinfo{author}{\bibfnamefont{E.~I.} \bibnamefont{Tiktopulo}},
  \bibinfo{author}{\bibfnamefont{V.~N.} \bibnamefont{Uversky}},
  \bibinfo{author}{\bibfnamefont{V.~B.} \bibnamefont{Lushchik}},
  \bibinfo{author}{\bibfnamefont{S.~I.} \bibnamefont{Klenin}},
  \bibinfo{author}{\bibfnamefont{V.~E.} \bibnamefont{Bychkova}},
  \bibnamefont{and} \bibinfo{author}{\bibfnamefont{O.~B.}
  \bibnamefont{Ptitsyn}}, \bibinfo{journal}{Macromolecules}
  \textbf{\bibinfo{volume}{28}}, \bibinfo{pages}{7519} (\bibinfo{year}{1995}).

\bibitem[{\citenamefont{Wu and Zhou}(1996)}]{WuZhou1996}
\bibinfo{author}{\bibfnamefont{C.}~\bibnamefont{Wu}} \bibnamefont{and}
  \bibinfo{author}{\bibfnamefont{S.}~\bibnamefont{Zhou}},
  \bibinfo{journal}{Phys. Rev. Lett.} \textbf{\bibinfo{volume}{77}},
  \bibinfo{pages}{3053} (\bibinfo{year}{1996}).

\bibitem[{\citenamefont{Wu and Wang}(1998)}]{WuWang1998}
\bibinfo{author}{\bibfnamefont{C.}~\bibnamefont{Wu}} \bibnamefont{and}
  \bibinfo{author}{\bibfnamefont{X.}~\bibnamefont{Wang}},
  \bibinfo{journal}{Phys. Rev. Lett.} \textbf{\bibinfo{volume}{80}},
  \bibinfo{pages}{4092} (\bibinfo{year}{1998}).

\bibitem[{\citenamefont{Hansen et~al.}(1998)\citenamefont{Hansen, Jensen,
  Sneppen, and Zocchi}}]{HansenJensenSneppenZocchi1998}
\bibinfo{author}{\bibfnamefont{A.}~\bibnamefont{Hansen}},
  \bibinfo{author}{\bibfnamefont{M.~H.} \bibnamefont{Jensen}},
  \bibinfo{author}{\bibfnamefont{K.}~\bibnamefont{Sneppen}}, \bibnamefont{and}
  \bibinfo{author}{\bibfnamefont{G.}~\bibnamefont{Zocchi}},
  \bibinfo{journal}{Eur. Phys. J. B} \textbf{\bibinfo{volume}{6}},
  \bibinfo{pages}{157} (\bibinfo{year}{1998}).

\bibitem[{\citenamefont{Hansen et~al.}(1999)\citenamefont{Hansen, Jensen,
  Sneppen, and Zocchi}}]{HansenJensenSneppenZocchi1999}
\bibinfo{author}{\bibfnamefont{A.}~\bibnamefont{Hansen}},
  \bibinfo{author}{\bibfnamefont{M.~H.} \bibnamefont{Jensen}},
  \bibinfo{author}{\bibfnamefont{K.}~\bibnamefont{Sneppen}}, \bibnamefont{and}
  \bibinfo{author}{\bibfnamefont{G.}~\bibnamefont{Zocchi}},
  \bibinfo{journal}{Eur. Phys. J. B} \textbf{\bibinfo{volume}{10}},
  \bibinfo{pages}{193} (\bibinfo{year}{1999}).

\bibitem[{\citenamefont{Bakk et~al.}(2001{\natexlab{a}})\citenamefont{Bakk,
  H{\o}ye, Hansen, and Sneppen}}]{BakkHoyeHansenSneppen2001}
\bibinfo{author}{\bibfnamefont{A.}~\bibnamefont{Bakk}},
  \bibinfo{author}{\bibfnamefont{J.~S.} \bibnamefont{H{\o}ye}},
  \bibinfo{author}{\bibfnamefont{A.}~\bibnamefont{Hansen}}, \bibnamefont{and}
  \bibinfo{author}{\bibfnamefont{K.}~\bibnamefont{Sneppen}},
  \bibinfo{journal}{J. Theor. Biol.} \textbf{\bibinfo{volume}{210}},
  \bibinfo{pages}{367} (\bibinfo{year}{2001}{\natexlab{a}}).

\bibitem[{\citenamefont{Bakk et~al.}(2001{\natexlab{b}})\citenamefont{Bakk,
  Hansen, and Sneppen}}]{BakkHansenSneppen2001}
\bibinfo{author}{\bibfnamefont{A.}~\bibnamefont{Bakk}},
  \bibinfo{author}{\bibfnamefont{A.}~\bibnamefont{Hansen}}, \bibnamefont{and}
  \bibinfo{author}{\bibfnamefont{K.}~\bibnamefont{Sneppen}},
  \bibinfo{journal}{Physica A} \textbf{\bibinfo{volume}{291}},
  \bibinfo{pages}{60} (\bibinfo{year}{2001}{\natexlab{b}}).

\bibitem[{\citenamefont{Rios and Caldarelli}(2000)}]{DelosriosCaldarelli2000}
\bibinfo{author}{\bibfnamefont{P.~D.~L.} \bibnamefont{Rios}} \bibnamefont{and}
  \bibinfo{author}{\bibfnamefont{G.}~\bibnamefont{Caldarelli}},
  \bibinfo{journal}{Phys. Rev. E} \textbf{\bibinfo{volume}{62}},
  \bibinfo{pages}{8449} (\bibinfo{year}{2000}).

\bibitem[{\citenamefont{Rios and Caldarelli}(2001)}]{DelosriosCaldarelli2001}
\bibinfo{author}{\bibfnamefont{P.~D.~L.} \bibnamefont{Rios}} \bibnamefont{and}
  \bibinfo{author}{\bibfnamefont{G.}~\bibnamefont{Caldarelli}},
  \bibinfo{journal}{Phys. Rev. E} \textbf{\bibinfo{volume}{63}},
  \bibinfo{pages}{031802} (\bibinfo{year}{2001}).

\bibitem[{\citenamefont{Bruscolini and
  Casetti}(2000)}]{BruscoliniCasetti2000pre}
\bibinfo{author}{\bibfnamefont{P.}~\bibnamefont{Bruscolini}} \bibnamefont{and}
  \bibinfo{author}{\bibfnamefont{L.}~\bibnamefont{Casetti}},
  \bibinfo{journal}{Phys. Rev. E} \textbf{\bibinfo{volume}{61}},
  \bibinfo{pages}{R2208} (\bibinfo{year}{2000}).

\bibitem[{\citenamefont{Bruscolini and
  Casetti}(2001)}]{BruscoliniCasetti2001pre}
\bibinfo{author}{\bibfnamefont{P.}~\bibnamefont{Bruscolini}} \bibnamefont{and}
  \bibinfo{author}{\bibfnamefont{L.}~\bibnamefont{Casetti}},
  \bibinfo{journal}{Phys. Rev. E} \textbf{\bibinfo{volume}{64}},
  \bibinfo{pages}{051805} (\bibinfo{year}{2001}).

\bibitem[{\citenamefont{Bruscolini et~al.}(2001)\citenamefont{Bruscolini,
  Buzano, Pelizzola, and Pretti}}]{BruscoliniBuzanoPelizzolaPretti2001}
\bibinfo{author}{\bibfnamefont{P.}~\bibnamefont{Bruscolini}},
  \bibinfo{author}{\bibfnamefont{C.}~\bibnamefont{Buzano}},
  \bibinfo{author}{\bibfnamefont{A.}~\bibnamefont{Pelizzola}},
  \bibnamefont{and} \bibinfo{author}{\bibfnamefont{M.}~\bibnamefont{Pretti}},
  \bibinfo{journal}{Phys. Rev. E} \textbf{\bibinfo{volume}{64}},
  \bibinfo{pages}{050801(R)} (\bibinfo{year}{2001}).

\bibitem[{\citenamefont{Bruscolini et~al.}(2002)\citenamefont{Bruscolini,
  Buzano, Pelizzola, and Pretti}}]{BruscoliniBuzanoPelizzolaPretti2002}
\bibinfo{author}{\bibfnamefont{P.}~\bibnamefont{Bruscolini}},
  \bibinfo{author}{\bibfnamefont{C.}~\bibnamefont{Buzano}},
  \bibinfo{author}{\bibfnamefont{A.}~\bibnamefont{Pelizzola}},
  \bibnamefont{and} \bibinfo{author}{\bibfnamefont{M.}~\bibnamefont{Pretti}},
  \bibinfo{journal}{Macromol. Symp.} \textbf{\bibinfo{volume}{181}},
  \bibinfo{pages}{261} (\bibinfo{year}{2002}).

\bibitem[{\citenamefont{Salvi and Rios}(2003)}]{SalviDelosrios2003}
\bibinfo{author}{\bibfnamefont{G.}~\bibnamefont{Salvi}} \bibnamefont{and}
  \bibinfo{author}{\bibfnamefont{P.~D.~L.} \bibnamefont{Rios}},
  \bibinfo{journal}{Phys. Rev. Lett.} \textbf{\bibinfo{volume}{91}},
  \bibinfo{pages}{258102} (\bibinfo{year}{2003}).

\bibitem[{\citenamefont{Paschek and Garc\`{\i}a}(2004)}]{PaschekGarcia2004}
\bibinfo{author}{\bibfnamefont{D.}~\bibnamefont{Paschek}} \bibnamefont{and}
  \bibinfo{author}{\bibfnamefont{A.~E.} \bibnamefont{Garc\`{\i}a}},
  \bibinfo{journal}{Phys. Rev. Lett.} \textbf{\bibinfo{volume}{93}},
  \bibinfo{pages}{238105} (\bibinfo{year}{2004}).

\bibitem[{\citenamefont{Polson and Zuckermann}(2000)}]{PolsonZuckermann2000}
\bibinfo{author}{\bibfnamefont{J.~M.} \bibnamefont{Polson}} \bibnamefont{and}
  \bibinfo{author}{\bibfnamefont{M.~J.} \bibnamefont{Zuckermann}},
  \bibinfo{journal}{J. Chem. Phys.} \textbf{\bibinfo{volume}{113}},
  \bibinfo{pages}{1283} (\bibinfo{year}{2000}).

\bibitem[{\citenamefont{Ghosh et~al.}(2005)\citenamefont{Ghosh, Kalra, and
  Garde}}]{GhoshKalraGarde2005}
\bibinfo{author}{\bibfnamefont{T.}~\bibnamefont{Ghosh}},
  \bibinfo{author}{\bibfnamefont{A.}~\bibnamefont{Kalra}}, \bibnamefont{and}
  \bibinfo{author}{\bibfnamefont{S.}~\bibnamefont{Garde}}, \bibinfo{journal}{J.
  Phys. Chem. B} \textbf{\bibinfo{volume}{109}}, \bibinfo{pages}{642}
  (\bibinfo{year}{2005}).

\bibitem[{\citenamefont{Paschek et~al.}(2005)\citenamefont{Paschek, Nonn, and
  Geiger}}]{PaschekNonnGeiger2005}
\bibinfo{author}{\bibfnamefont{D.}~\bibnamefont{Paschek}},
  \bibinfo{author}{\bibfnamefont{S.}~\bibnamefont{Nonn}}, \bibnamefont{and}
  \bibinfo{author}{\bibfnamefont{A.}~\bibnamefont{Geiger}},
  \bibinfo{journal}{Phys. Chem. Chem. Phys.} \textbf{\bibinfo{volume}{7}},
  \bibinfo{pages}{2780} (\bibinfo{year}{2005}).

\bibitem[{\citenamefont{Pretti and Buzano}(2004)}]{PrettiBuzano2004}
\bibinfo{author}{\bibfnamefont{M.}~\bibnamefont{Pretti}} \bibnamefont{and}
  \bibinfo{author}{\bibfnamefont{C.}~\bibnamefont{Buzano}},
  \bibinfo{journal}{J. Chem. Phys.} \textbf{\bibinfo{volume}{121}},
  \bibinfo{pages}{11856} (\bibinfo{year}{2004}).

\bibitem[{\citenamefont{Pretti and Buzano}(2005)}]{PrettiBuzano2005}
\bibinfo{author}{\bibfnamefont{M.}~\bibnamefont{Pretti}} \bibnamefont{and}
  \bibinfo{author}{\bibfnamefont{C.}~\bibnamefont{Buzano}},
  \bibinfo{journal}{J. Chem. Phys.} \textbf{\bibinfo{volume}{123}},
  \bibinfo{pages}{024506} (\bibinfo{year}{2005}).

\bibitem[{\citenamefont{Roberts and
  Debenedetti}(1996)}]{RobertsDebenedetti1996}
\bibinfo{author}{\bibfnamefont{C.~J.} \bibnamefont{Roberts}} \bibnamefont{and}
  \bibinfo{author}{\bibfnamefont{P.~G.} \bibnamefont{Debenedetti}},
  \bibinfo{journal}{J. Chem. Phys.} \textbf{\bibinfo{volume}{105}},
  \bibinfo{pages}{658} (\bibinfo{year}{1996}).

\bibitem[{\citenamefont{Roberts et~al.}(1996)\citenamefont{Roberts,
  Panagiotopoulos, and Debenedetti}}]{RobertsPanagiotopoulosDebenedetti1996}
\bibinfo{author}{\bibfnamefont{C.~J.} \bibnamefont{Roberts}},
  \bibinfo{author}{\bibfnamefont{A.~Z.} \bibnamefont{Panagiotopoulos}},
  \bibnamefont{and} \bibinfo{author}{\bibfnamefont{P.~G.}
  \bibnamefont{Debenedetti}}, \bibinfo{journal}{Phys. Rev. Lett.}
  \textbf{\bibinfo{volume}{77}}, \bibinfo{pages}{4386} (\bibinfo{year}{1996}).

\bibitem[{\citenamefont{Roberts et~al.}(1998)\citenamefont{Roberts,
  Karayiannakis, and Debenedetti}}]{RobertsKarayiannakisDebenedetti1998}
\bibinfo{author}{\bibfnamefont{C.~J.} \bibnamefont{Roberts}},
  \bibinfo{author}{\bibfnamefont{G.~A.} \bibnamefont{Karayiannakis}},
  \bibnamefont{and} \bibinfo{author}{\bibfnamefont{P.~G.}
  \bibnamefont{Debenedetti}}, \bibinfo{journal}{Ind. Eng. Chem. Res.}
  \textbf{\bibinfo{volume}{37}}, \bibinfo{pages}{3012} (\bibinfo{year}{1998}).

\bibitem[{\citenamefont{Pretti}(2003)}]{Pretti2003}
\bibinfo{author}{\bibfnamefont{M.}~\bibnamefont{Pretti}}, \bibinfo{journal}{J.
  Stat. Phys.} \textbf{\bibinfo{volume}{111}}, \bibinfo{pages}{993}
  (\bibinfo{year}{2003}).

\bibitem[{\citenamefont{Pretti}(2002)}]{Pretti2002}
\bibinfo{author}{\bibfnamefont{M.}~\bibnamefont{Pretti}},
  \bibinfo{journal}{Phys. Rev. E} \textbf{\bibinfo{volume}{66}},
  \bibinfo{pages}{061802} (\bibinfo{year}{2002}).

\bibitem[{\citenamefont{Mishima and Stanley}(1998)}]{MishimaStanley1998}
\bibinfo{author}{\bibfnamefont{O.}~\bibnamefont{Mishima}} \bibnamefont{and}
  \bibinfo{author}{\bibfnamefont{H.~E.} \bibnamefont{Stanley}},
  \bibinfo{journal}{Nature} \textbf{\bibinfo{volume}{392}},
  \bibinfo{pages}{164} (\bibinfo{year}{1998}).

\bibitem[{\citenamefont{Speedy}(1982)}]{Speedy1982I}
\bibinfo{author}{\bibfnamefont{R.~J.} \bibnamefont{Speedy}},
  \bibinfo{journal}{J. Phys. Chem.} \textbf{\bibinfo{volume}{86}},
  \bibinfo{pages}{982} (\bibinfo{year}{1982}).

\bibitem[{\citenamefont{Zheng et~al.}(1991)\citenamefont{Zheng, Durben, Wolf,
  and Angell}}]{ZhengDurbenWolfAngell1991}
\bibinfo{author}{\bibfnamefont{Q.}~\bibnamefont{Zheng}},
  \bibinfo{author}{\bibfnamefont{D.~J.} \bibnamefont{Durben}},
  \bibinfo{author}{\bibfnamefont{G.~H.} \bibnamefont{Wolf}}, \bibnamefont{and}
  \bibinfo{author}{\bibfnamefont{C.~A.} \bibnamefont{Angell}},
  \bibinfo{journal}{Science} \textbf{\bibinfo{volume}{254}},
  \bibinfo{pages}{829} (\bibinfo{year}{1991}).

\bibitem[{\citenamefont{Speedy}(2004)}]{Speedy2004}
\bibinfo{author}{\bibfnamefont{R.}~\bibnamefont{Speedy}}, \bibinfo{journal}{J.
  Phys.: Condens. Matter} \textbf{\bibinfo{volume}{16}}, \bibinfo{pages}{6811}
  (\bibinfo{year}{2004}).

\bibitem[{\citenamefont{Debenedetti}(2004)}]{Debenedetti2004}
\bibinfo{author}{\bibfnamefont{P.~G.} \bibnamefont{Debenedetti}},
  \bibinfo{journal}{J. Phys.: Condens. Matter} \textbf{\bibinfo{volume}{16}},
  \bibinfo{pages}{6815} (\bibinfo{year}{2004}).

\bibitem[{\citenamefont{Vanderzande}(1998)}]{Vanderzande1998}
\bibinfo{author}{\bibfnamefont{C.}~\bibnamefont{Vanderzande}},
  \emph{\bibinfo{title}{Lattice Models of Polymers}}
  (\bibinfo{publisher}{Cambridge University Press, Cambridge},
  \bibinfo{year}{1998}).

\bibitem[{\citenamefont{Panick et~al.}(1999)\citenamefont{Panick, Vidugiris,
  Malessa, Rapp, Winter, and Royer}}]{Panick_et_al1999}
\bibinfo{author}{\bibfnamefont{G.}~\bibnamefont{Panick}},
  \bibinfo{author}{\bibfnamefont{G.~J.~A.} \bibnamefont{Vidugiris}},
  \bibinfo{author}{\bibfnamefont{R.}~\bibnamefont{Malessa}},
  \bibinfo{author}{\bibfnamefont{G.}~\bibnamefont{Rapp}},
  \bibinfo{author}{\bibfnamefont{R.}~\bibnamefont{Winter}}, \bibnamefont{and}
  \bibinfo{author}{\bibfnamefont{C.~A.} \bibnamefont{Royer}},
  \bibinfo{journal}{Biochemistry} \textbf{\bibinfo{volume}{38}},
  \bibinfo{pages}{4157} (\bibinfo{year}{1999}).

\bibitem[{\citenamefont{Herberhold and Winter}(2002)}]{HerberholdWinter2002}
\bibinfo{author}{\bibfnamefont{H.}~\bibnamefont{Herberhold}} \bibnamefont{and}
  \bibinfo{author}{\bibfnamefont{R.}~\bibnamefont{Winter}},
  \bibinfo{journal}{Biochemistry} \textbf{\bibinfo{volume}{41}},
  \bibinfo{pages}{2396} (\bibinfo{year}{2002}).

\end{thebibliography}

\end{document}